\documentclass[runningheads]{llncs}
\usepackage[T1]{fontenc}

\usepackage{import}
\usepackage{algorithm}
\usepackage{algorithmicx}
\usepackage[noend]{algpseudocode}

\algnewcommand{\algorithmicswitch}{\textbf{switch}}
\algdef{SE}[SWITCH]{Switch}{EndSwitch}[1]{\algorithmicswitch\ #1\ \algorithmicdo}{\algorithmicend\ \algorithmicswitch}%
\algtext*{EndSwitch}%

\algnewcommand{\algorithmiccase}{\textbf{case}}
\algdef{SE}[CASE]{Case}{EndCase}[1]{\algorithmiccase\ #1}{\algorithmicend\ \algorithmiccase}%
\algtext*{EndCase}%

\algnewcommand{\algorithmicon}{\textbf{on}}
\algdef{SE}[ON]{On}{EndOn}[1]{\algorithmicon\ #1\ \algorithmicdo}{\algorithmicend\ \algorithmicon}%
\algtext*{EndOn}%

\algnewcommand{\algorithmicat}{\textbf{at}}
\algdef{SE}[AT]{At}{EndAt}[1]{\algorithmicat\ #1\ \algorithmicdo}{\algorithmicend\ \algorithmicat}%
\algtext*{EndAt}%

\algnewcommand{\algorithmicrealfunction}{\textbf{function}}
\algdef{SE}[REALFUNCTION]{RealFunction}{EndRealFunction}[1]{\algorithmicrealfunction\ #1\ \algorithmicdo}{\algorithmicend\ \algorithmicrealfunction}%
\algtext*{EndRealFunction}%

\algnewcommand{\algorithmicthroughout}{\textbf{do throughout}}
\algdef{SE}[Throughout]{Throughout}{EndThroughout}[1]{\algorithmicthroughout\ #1\ \algorithmicdo}{\algorithmicend\ \algorithmicthroughout}%
\algtext*{EndThroughout}%

\algrenewcommand{\algorithmicdo}{}
\algrenewcommand{\algorithmicthen}{}

\algnewcommand{\algorithmicgoto}{\textbf{goto}}%
\algnewcommand{\Goto}[1]{\algorithmicgoto~\ref{#1}}%

\algnewcommand{\algorithmicassert}{\textbf{assert}}%
\algnewcommand{\Assert}[1]{\algorithmicassert~{#1}}%

\algnewcommand{\algorithmicbreak}{\textbf{break}}%
\algnewcommand{\Break}[0]{\algorithmicbreak}%

\algnewcommand{\algorithmicwaiton}{\textbf{wait on}}%
\algnewcommand{\WaitOn}[1]{\algorithmicwaiton~{#1}}%

\algnewcommand{\LineComment}[1]{\State \(\triangleright\) \textit{#1}}
\algnewcommand{\InlineRequire}[1]{\textbf{require} {#1}}

\usepackage[utf8]{inputenc}

\PassOptionsToPackage{hyphens}{url}%
\usepackage{hyperref}
\usepackage{graphicx}

\usepackage{amsmath}
\usepackage{amssymb}
\usepackage{amsfonts}
\usepackage{mathtools}

\usepackage{shortsym}

\allowdisplaybreaks

\usepackage[shortlabels,inline]{enumitem}
\setlist[itemize]{leftmargin=1.25em}
\setlist[enumerate]{leftmargin=1.75em}

\usepackage{xstring}   %

\usepackage{subfig}

\usepackage{breakcites}

\usepackage{import}
\makeatletter
\renewcommand*\env@matrix[1][*\c@MaxMatrixCols c]{%
    \hskip -\arraycolsep
    \let\@ifnextchar\new@ifnextchar
    \array{#1}}
\makeatother

\DeclarePairedDelimiter\abs{\lvert}{\rvert}
\DeclarePairedDelimiter\len{\lvert}{\rvert}
\DeclarePairedDelimiter\norm{\lVert}{\rVert}

\makeatletter
\let\oldabs\abs
\def\abs{\@ifstar{\oldabs}{\oldabs*}}
\let\oldlen\len
\def\len{\@ifstar{\oldlen}{\oldlen*}}
\let\oldnorm\norm
\def\norm{\@ifstar{\oldnorm}{\oldnorm*}}
\makeatother

\usepackage{pifont}

\usepackage{dsfont}

\usepackage{xspace}

\newcommand{\ie}[0]{\emph{i.e.}\xspace}
\newcommand{\eg}[0]{\emph{e.g.}\xspace}

\usepackage{xcolor}

\definecolor{jnSUCardinalRed}{HTML}{8c1515}
\definecolor{jnSUCardinalRedLight}{HTML}{B83A4B}
\definecolor{jnSUCardinalRedDark}{HTML}{820000}
\definecolor{jnSUWhite}{HTML}{ffffff}
\definecolor{jnSUCoolGrey}{HTML}{53565A}
\definecolor{jnSUBlack}{HTML}{2e2d29}
\definecolor{jnSUBlack100}{HTML}{2e2d29}
\definecolor{jnSUBlack90}{HTML}{43423E}
\definecolor{jnSUBlack80}{HTML}{585754}
\definecolor{jnSUBlack70}{HTML}{6D6C69}
\definecolor{jnSUBlack60}{HTML}{767674}
\definecolor{jnSUBlack50}{HTML}{979694}
\definecolor{jnSUBlack40}{HTML}{ABABA9}
\definecolor{jnSUBlack30}{HTML}{C0C0BF}
\definecolor{jnSUBlack20}{HTML}{D5D5D4}
\definecolor{jnSUBlack10}{HTML}{EAEAEA}

\definecolor{jnSUPaloAlto}{HTML}{175E54}
\definecolor{jnSUPaloAltoLight}{HTML}{2D716F}
\definecolor{jnSUPaloAltoDark}{HTML}{014240}
\definecolor{jnSUPaloVerde}{HTML}{279989}
\definecolor{jnSUPaloVerdeLight}{HTML}{59B3A9}
\definecolor{jnSUPaloVerdeDark}{HTML}{017E7C}
\definecolor{jnSUOlive}{HTML}{8F993E}
\definecolor{jnSUOliveLight}{HTML}{A6B168}
\definecolor{jnSUOliveDark}{HTML}{7A863B}
\definecolor{jnSUBay}{HTML}{6FA287}
\definecolor{jnSUBayLight}{HTML}{8AB8A7}
\definecolor{jnSUBayDark}{HTML}{417865}
\definecolor{jnSUSky}{HTML}{4298B5}
\definecolor{jnSUSkyLight}{HTML}{67AFD2}
\definecolor{jnSUSkyDark}{HTML}{016895}
\definecolor{jnSULagunita}{HTML}{007C92}
\definecolor{jnSULagunitaLight}{HTML}{009AB4}
\definecolor{jnSULagunitaDark}{HTML}{006B81}
\definecolor{jnSUPoppy}{HTML}{E98300}
\definecolor{jnSUPoppyLight}{HTML}{F9A44A}
\definecolor{jnSUPoppyDark}{HTML}{D1660F}
\definecolor{jnSUSpirited}{HTML}{E04F39}
\definecolor{jnSUSpiritedLight}{HTML}{F4795B}
\definecolor{jnSUSpiritedDark}{HTML}{C74632}
\definecolor{jnSUIlluminating}{HTML}{FEDD5C}
\definecolor{jnSUIlluminatingLight}{HTML}{FFE781}
\definecolor{jnSUIlluminatingDark}{HTML}{FEC51D}
\definecolor{jnSUPlum}{HTML}{620059}
\definecolor{jnSUPlumLight}{HTML}{734675}
\definecolor{jnSUPlumDark}{HTML}{350D36}
\definecolor{jnSUBrick}{HTML}{651C32}
\definecolor{jnSUBrickLight}{HTML}{7F2D48}
\definecolor{jnSUBrickDark}{HTML}{42081B}
\definecolor{jnSUArchway}{HTML}{5D4B3C}
\definecolor{jnSUArchwayLight}{HTML}{766253}
\definecolor{jnSUArchwayDark}{HTML}{2F2424}
\definecolor{jnSUStone}{HTML}{7F7776}
\definecolor{jnSUStoneLight}{HTML}{D4D1D1}
\definecolor{jnSUStoneDark}{HTML}{544948}
\definecolor{jnSUFog}{HTML}{DAD7CB}
\definecolor{jnSUFogLight}{HTML}{F4F4F4}
\definecolor{jnSUFogDark}{HTML}{B6B1A9}

\definecolor{jnSUDigitalRed}{HTML}{B1040E}
\definecolor{jnSUDigitalRedLight}{HTML}{E50808}
\definecolor{jnSUDigitalRedDark}{HTML}{820000}
\definecolor{jnSUDigitalBlue}{HTML}{006CB8}
\definecolor{jnSUDigitalBlueLight}{HTML}{6FC3FF}
\definecolor{jnSUDigitalBlueDark}{HTML}{00548f}
\definecolor{jnSUDigitalGreen}{HTML}{008566}
\definecolor{jnSUDigitalGreenLight}{HTML}{1AECBA}
\definecolor{jnSUDigitalGreenDark}{HTML}{006F54}

\definecolor{myParula01Blue}{RGB}{0,114,189}
\definecolor{myParula02Orange}{RGB}{217,83,25}
\definecolor{myParula03Yellow}{RGB}{237,177,32}
\definecolor{myParula04Purple}{RGB}{126,47,142}
\definecolor{myParula05Green}{RGB}{119,172,48}
\definecolor{myParula06LightBlue}{RGB}{77,190,238}
\definecolor{myParula07Red}{RGB}{162,20,47}

\usepackage{tikz}
\usetikzlibrary{calc}
\usetikzlibrary{arrows}
\usetikzlibrary{arrows.meta}
\usetikzlibrary{patterns}
\usetikzlibrary{positioning}
\usetikzlibrary{decorations.pathreplacing}
\usetikzlibrary{calligraphy}
\usetikzlibrary{shapes.misc}
\usetikzlibrary{shapes.symbols}
\usetikzlibrary{shapes.arrows}
\usetikzlibrary{spy}
\usetikzlibrary{shapes.geometric}
\usetikzlibrary{intersections}
\usetikzlibrary{fit}

\usepackage{pgfplots}
\pgfplotsset{compat=1.17}
\usepgfplotslibrary{fillbetween}

\pgfplotsset{
    discard if not/.style 2 args={
            x filter/.code={
                    \edef\tempa{\thisrow{#1}}
                    \edef\tempb{#2}
                    \ifx\tempa\tempb
                    \else
                        
                    \fi
                }
        },
}

\tikzset{myparula11/.style={color=myParula01Blue,solid,mark=+,mark options={solid}}}
\tikzset{myparula12/.style={color=myParula01Blue,densely dashed,mark=x,mark options={solid}}}
\tikzset{myparula13/.style={color=myParula01Blue,densely dotted,mark=o,mark options={solid}}}
\tikzset{myparula14/.style={color=myParula01Blue,dashdotted,mark=triangle,mark options={solid}}}
\tikzset{myparula15/.style={color=myParula01Blue,dashdotdotted,mark=square,mark options={solid}}}

\tikzset{myparula21/.style={color=myParula02Orange,solid,mark=+,mark options={solid}}}
\tikzset{myparula22/.style={color=myParula02Orange,densely dashed,mark=x,mark options={solid}}}
\tikzset{myparula23/.style={color=myParula02Orange,densely dotted,mark=o,mark options={solid}}}
\tikzset{myparula24/.style={color=myParula02Orange,dashdotted,mark=triangle,mark options={solid}}}
\tikzset{myparula25/.style={color=myParula02Orange,dashdotdotted,mark=square,mark options={solid}}}

\tikzset{myparula31/.style={color=myParula03Yellow,solid,mark=+,mark options={solid}}}
\tikzset{myparula32/.style={color=myParula03Yellow,densely dashed,mark=x,mark options={solid}}}
\tikzset{myparula33/.style={color=myParula03Yellow,densely dotted,mark=o,mark options={solid}}}
\tikzset{myparula34/.style={color=myParula03Yellow,dashdotted,mark=triangle,mark options={solid}}}
\tikzset{myparula35/.style={color=myParula03Yellow,dashdotdotted,mark=square,mark options={solid}}}

\tikzset{myparula41/.style={color=myParula04Purple,solid,mark=+,mark options={solid}}}
\tikzset{myparula42/.style={color=myParula04Purple,densely dashed,mark=x,mark options={solid}}}
\tikzset{myparula43/.style={color=myParula04Purple,densely dotted,mark=o,mark options={solid}}}
\tikzset{myparula44/.style={color=myParula04Purple,dashdotted,mark=triangle,mark options={solid}}}
\tikzset{myparula45/.style={color=myParula04Purple,dashdotdotted,mark=square,mark options={solid}}}

\tikzset{myparula51/.style={color=myParula05Green,solid,mark=+,mark options={solid}}}
\tikzset{myparula52/.style={color=myParula05Green,densely dashed,mark=x,mark options={solid}}}
\tikzset{myparula53/.style={color=myParula05Green,densely dotted,mark=o,mark options={solid}}}
\tikzset{myparula54/.style={color=myParula05Green,dashdotted,mark=triangle,mark options={solid}}}
\tikzset{myparula55/.style={color=myParula05Green,dashdotdotted,mark=square,mark options={solid}}}

\tikzset{myparula61/.style={color=myParula06LightBlue,solid,mark=+,mark options={solid}}}
\tikzset{myparula62/.style={color=myParula06LightBlue,densely dashed,mark=x,mark options={solid}}}
\tikzset{myparula63/.style={color=myParula06LightBlue,densely dotted,mark=o,mark options={solid}}}
\tikzset{myparula64/.style={color=myParula06LightBlue,dashdotted,mark=triangle,mark options={solid}}}
\tikzset{myparula65/.style={color=myParula06LightBlue,dashdotdotted,mark=square,mark options={solid}}}

\tikzset{myparula71/.style={color=myParula07Red,solid,mark=+,mark options={solid}}}
\tikzset{myparula72/.style={color=myParula07Red,densely dashed,mark=x,mark options={solid}}}
\tikzset{myparula73/.style={color=myParula07Red,densely dotted,mark=o,mark options={solid}}}
\tikzset{myparula74/.style={color=myParula07Red,dashdotted,mark=triangle,mark options={solid}}}
\tikzset{myparula75/.style={color=myParula07Red,dashdotdotted,mark=square,mark options={solid}}}

\pgfplotsset{
    mysimpleplot/.style = {
            every axis plot/.prefix style={thick},
            width=1.05\linewidth,
            height=0.75\linewidth,
            title style={font=\footnotesize,align=center},
            legend cell align=left,
            legend style={font=\footnotesize},
            legend columns=3,
            legend style={
                    at={(0.5,1)},
                    yshift=0.3em,
                    anchor=south,
                    draw=none,
                    /tikz/every even column/.append style={
                            column sep=0.3em
                        },
                    cells={
                            align=left
                        }
                },
            grid=both,
            minor tick num=4,
            major grid style={solid,draw=gray!50},
            minor grid style={densely dotted,draw=gray!50},
            label style={font=\footnotesize,align=center},
            tick label style={font=\footnotesize},
        },
}

\pgfplotsset{
    myresultsplot01/.style={
            legend style={
                    at={(0.5,1.15)},
                    anchor=north,
                    legend columns=-1,
                    draw=none,
                    /tikz/every even column/.append style={column sep=1em},
                    cells={align=left},
                    font=\small,
                },
            enlarge x limits=0.15,
            ybar,
            bar width=6mm,
            xtick=data,
            width=\linewidth,
            height=0.6\linewidth,
            ymin=0,
            grid=both,
            minor y tick num=4,
            minor x tick num=1,
            major grid style={solid,draw=gray!50},
            minor grid style={densely dotted,draw=gray!50},
            label style={font=\small},
            tick label style={font=\small},
        },
}

\pgfplotsset{
    mysimpleresilienceplot01/.style = {
            mysimpleplot,
            ylabel={Adversary resilience $\beta$},
            height=0.45\linewidth,
            width=\linewidth,
            ymin=0,ymax=0.5,
            ytick={0,0.1,0.2,0.3,0.4,0.5},
            xmin=1e-5,xmax=1e2,
            grid=major,
        },
}

\pgfplotsset{
    mybandwidthplot01/.style = {
            mysimpleplot,
            ylabel={Minimum bandwidth $C$ (Mbps)},
            xlabel={Adversary resilience $\beta$},
            height=0.6\linewidth,
            width=\linewidth,
            xmin=0,xmax=0.5,
            xtick={0,0.1,0.2,0.3,0.4,0.5},
        },
}

\tikzset{blockchainold/.style={
            x=1.5cm,
            y=0.6cm,
            node distance=0.5cm,
            block/.style = {
                    minimum width=0.25cm,
                    minimum height=0.25cm,
                    draw,
                    shade,
                    top color=white,
                    bottom color=black!10,
                },
            block-adv1/.style = {
                    block,
                    bottom color=myParula01Blue!50,
                    draw=myParula01Blue!50!black
                },
            block-adv2/.style = {
                    block,
                    bottom color=myParula07Red!50,
                    draw=myParula07Red!50!black,
                },
            block-adv3/.style = {
                    block,
                    bottom color=myParula05Green!50,
                    draw=myParula05Green!50!black,
                },
            block-green/.style = {
                    block,
                    bottom color=myParula05Green!50,
                    draw=myParula05Green!50!black,
                },
            block-red/.style = {
                    block,
                    bottom color=myParula07Red!50,
                    draw=myParula07Red!70!black,
                },
            block-gray/.style = {
                    block,
                    bottom color=black!30,
                },
            block-big/.style = {
                    minimum width=0.7cm,
                    minimum height=0.7cm,
                    draw,
                    shade,
                    top color=white,
                    bottom color=black!10,
                },
            branch/.style = {
                    minimum width=0.1cm,
                    minimum height=0.1cm,
                    draw,
                    circle,
                    inner sep=0,
                    fill=black,
                },
            link/.style = {
                    -latex,
                },
            hiddenlink/.style = {
                    dashed,
                },
            hiddenlink-adv1/.style = {
                    hiddenlink,
                    draw=myParula01Blue!50!black,
                },
            hiddenlink-adv2/.style = {
                    hiddenlink,
                    draw=myParula07Red!50!black,
                },
            hiddenlink-adv3/.style = {
                    hiddenlink,
                    draw=myParula05Green!50!black,
                },
            label/.style = {
                },
            label-adv1/.style = {
                    label,
                    text=myParula01Blue!50!black,
                },
            label-adv2/.style = {
                    label,
                    text=myParula07Red!50!black,
                },
            label-adv3/.style = {
                    label,
                    text=myParula05Green!50!black,
                },
        }
}

\tikzset{blockchain/.style={
            x=0.5cm,
            y=0.55cm,
            node distance=0.5cm,
            block/.style = {
                    minimum width=0.3cm,
                    minimum height=0.3cm,
                    draw,
                    shade,
                    top color=white,
                    bottom color=black!10,
                    inner sep=0,
                },
            block-adv/.style = {
                    block,
                    bottom color=myParula07Red!50,
                    draw=myParula07Red!50!black,
                },
            block-hon/.style = {
                    block,
                    bottom color=myParula05Green!50,
                    draw=myParula05Green!50!black,
                },
            block-blank/.style = {
                    minimum width=0.3cm,
                    minimum height=0.3cm,
                    rounded corners,
                    inner sep=0,
                },
            link/.style = {
                    -latex,
                },
            link-adv/.style = {
                    link,
                },
            link-hon/.style = {
                    link,
                },
        }
}

\usepackage{multirow,booktabs}
\usepackage{makecell}

\usepackage{refstyle}

\usepackage[sort&compress,capitalize,nameinlink]{cleveref}

\crefname{figure}{Fig.}{Figs.}
\Crefname{figure}{Fig.}{Figs.}

\crefname{table}{Tab.}{Tabs.}
\Crefname{table}{Tab.}{Tabs.}

\crefname{section}{Sec.}{Secs.}
\Crefname{section}{Sec.}{Secs.}
\crefname{appendix}{App.}{Apps.}
\Crefname{appendix}{App.}{Apps.}

\crefname{algorithm}{Alg.}{Algs.}
\Crefname{algorithm}{Alg.}{Algs.}
\crefname{line}{l.}{ll.}
\Crefname{line}{l.}{ll.}

\crefname{proposition}{Prop.}{Props.}
\Crefname{proposition}{Prop.}{Props.}
\crefname{lemma}{Lem.}{Lems.}
\Crefname{lemma}{Lem.}{Lems.}
\crefname{theorem}{Thm.}{Thms.}
\Crefname{theorem}{Thm.}{Thms.}
\crefname{corollary}{Cor.}{Cors.}
\Crefname{corollary}{Cor.}{Cors.}
\crefname{definition}{Def.}{Defs.}
\Crefname{definition}{Def.}{Defs.}

\usepackage{color}

\usepackage{xspace}
\usepackage{pdflscape}
\usepackage{dialogue}

\newboolean{showcomments}
\setboolean{showcomments}{true}

\ifthenelse{\boolean{showcomments}}
{ \newcommand{\mynote}[3]{
    \protect\fbox{\bfseries\sffamily\scriptsize#1}
    {\small\textsf{\emph{\color{#3}{#2}}}}}}
{ \newcommand{\mynote}[3]{}}

\newcommand{\LOG}[0]{\ensuremath{L}}
\newcommand{\LOGi}[2]{\ensuremath{\pmb{L}_{#1}^{#2}}}

\newcommand{\consistent}{\ensuremath{\sim}}
\newcommand{\inconsistent}{\ensuremath{\not\sim}}

\newcommand{\LOGrec}{\ensuremath{\LOG_{\mathrm{rec}}}}

\newcommand{\LOGacki}[2]{\ensuremath{\smash{\widetilde{\pmb{L}}}_{#1}^{#2}}}
\newcommand{\LOGfinali}[2]{\ensuremath{\smash{\widehat{\pmb{L}}}_{#1}^{#2}}}

\newcommand{\tx}[0]{\ensuremath{\mathsf{tx}}}
\newcommand{\txs}[0]{\ensuremath{\mathsf{txs}}}

\newcommand{\valset}[0]{\ensuremath{\mathcal{N}}}
\newcommand{\valsetnew}[0]{\ensuremath{\mathcal{N}'}}

\newcommand{\genesis}[0]{\ensuremath{\LOG_{\mathrm{genesis}}}}

\newcommand{\tL}{\ensuremath{t^{\mathrm{L}}}}
\newcommand{\tS}{\ensuremath{t^{\mathrm{S}}}}

\newcommand{\rrec}{\ensuremath{r_{\mathrm{rec}}}}
\newcommand{\rmaj}{\ensuremath{r_{\mathrm{maj}}}}

\newcommand{\rboot}{\ensuremath{r^{\mathrm{wak}}}}
\newcommand{\PI}{\ensuremath{\widehat{\Pi}}}
\newcommand{\Rhrzn}{\ensuremath{R}}
\newcommand{\urec}{\ensuremath{u_{\mathrm{rec}}}}

\newcommand{\transcribe}{\ensuremath{\mathcal{W}}}
\newcommand{\untranscribe}{\ensuremath{\mathcal{C}}}
\newcommand{\transcript}{%
  \ifmmode%
    w%
  \else%
    witness\xspace %
  \fi%
}
\newcommand{\transcripts}{witnesses\xspace}

\newcommand{\bclatency}{\ensuremath{u_{\mathrm{BC}}}}

\newcommand{\msgSet}[0]{\ensuremath{\CS}}
\newcommand{\msgSetRec}[0]{\ensuremath{\CS_{\mathrm{rec}}}}

\newcommand{\clientcode}[1]{#1}
\newcommand{\valcode}[1]{#1}

\newcommand*\DiagramStep[1]{\tikz[baseline=(char.south)]{
                \node[shape=circle,draw=none,fill=black,text=white,inner sep=0.75pt] (char) {\tiny\textbf{#1}}; }}

\newcommand\tikzmark[1]{%
  \tikz[overlay,remember picture] \coordinate (#1);}

\begin{document}

\title{Better Safe than Sorry:\\Recovering after Adversarial Majority}

\author{Srivatsan Sridhar \and Dionysis Zindros \and David Tse}
\institute{Stanford University\\
\email{\{svatsan,dionyziz,dntse\}@stanford.edu}}

\maketitle
\begin{abstract}
The security of blockchain protocols is a combination of two properties: safety and liveness.
It is well known that no blockchain protocol can provide 
\emph{both} %
to 
sleepy (intermittently online)
clients under adversarial majority.
However, safety is more critical in that a single safety violation can cause users to lose money.
At the same time, liveness must not be lost forever.
We show that, in a synchronous network, it is possible to maintain \emph{safety} for all clients even during adversarial majority, \emph{and recover liveness} after honest majority is restored.
Our solution takes the form of a \emph{recovery gadget} that can be applied to any protocol with certificates
(such as HotStuff, Streamlet, Tendermint, and their variants).

\end{abstract}

\section{Introduction}
\label{sec:intro}

Eve the Evil Adversary is at it again. She has somehow managed to capture the \emph{majority} of the stake on our favorite proof-of-stake blockchain. She's about to double-spend: In one transaction, she sends multiplujillion dollars' worth of native tokens to one exchange owned by Alice. In another transaction, she sends the same amount to a different exchange, owned by Bob, spending the exact same money. 
Using her adversarial majority of voting power, she causes the chain to split into two forks. The first fork causes Alice to confirm the first transaction. The second fork causes Bob to confirm the second transaction. Before the exchanges realize what has transpired, Eve has withdrawn the money from both. As soon as the news hits, the whole community stops transacting: They convene at the bar to analyze what happened and which of the two forks to follow. They slash the misbehaving adversary, but a fight erupts between Alice and Bob: Neither of the two conflicting fork choices can make \emph{both} of them happy.

Here's a simple idea to avoid such a devastating scenario:
Alice and Bob
re-broadcast any chains they receive from the network, and so if conflicts occur, they are observed by both exchanges. In case of conflict, they each \emph{freeze} their operation and await human intervention. If such a deterring strategy is carefully followed, all honest participants, even those who joined late, always confirm ledgers consistent with one another, retaining safety even under adversarial majority. This simple trick achieves something remarkable: Our protocol lost liveness during adversarial majority but retains safety even up to a $99\%$ adversary, as long as everyone (including clients like Alice and Bob) gossips evidence of conflict and slightly delays confirming ledgers. Retaining safety without liveness makes sense: \emph{safety is more important than liveness}.

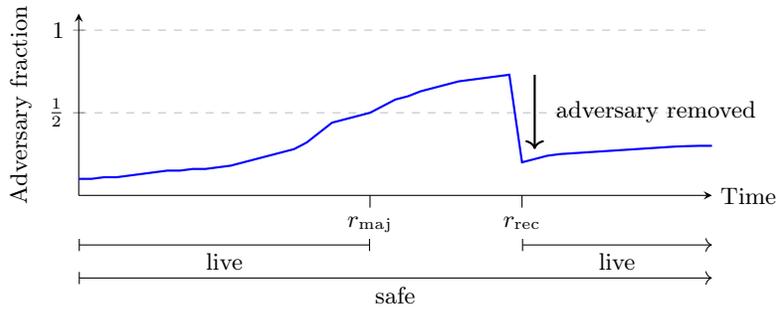
\begin{figure}[tb]
    \centering
    \begin{tikzpicture}
        \begin{axis}[
            height=4cm,
            width=10cm,
            axis lines=left,
            xlabel={Time},
            ylabel={Adversary fraction},
            ymin=0, ymax=1.1,
            xmin=0, xmax=50,
            xtick=\empty, ytick=\empty,
            ytick={0.5,1},
            yticklabels={\(\frac{1}{2}\), $1$},
            ymajorgrids=true,
            grid style=dashed,
            enlargelimits=false,
            clip=false,
            xlabel style={at={(axis description cs:1,0)}, anchor=west},
            x axis line style={-stealth}
        ]

        \addplot[blue,thick] coordinates {
        (0,0.1)  (1,0.1)  (2,0.11) (3,0.11) (4,0.12) 
        (5,0.13) (6,0.14) (7,0.15) (8,0.15) (9,0.16) 
        (10,0.16)(11,0.17)(12,0.18)(13,0.2) (14,0.22)
        (15,0.24)(16,0.26)(17,0.28)(18,0.32)(19,0.38)
        (20,0.44)(21,0.46)(22,0.48)(23,0.5) (24,0.54)
        (25,0.58)(26,0.6) (27,0.63)(28,0.65)(29,0.67)
        (30,0.69)(31,0.7) (32,0.71)(33,0.72)(34,0.73)
        (35,0.2) (36,0.22)(37,0.24)(38,0.25)(39,0.255)
        (40,0.26)(41,0.265)(42,0.27)(43,0.275)(44,0.28)
        (45,0.285)(46,0.29)(47,0.295)(48,0.298)(49,0.3)
        (50,0.3)
    };

        \coordinate (rmaj) at (axis cs:23,0);
        \coordinate (rrec) at (axis cs:35,0);
        \draw (rmaj) -- ++(0,-0.5em) node[below] {\( \rmaj \)};
        \draw (rrec) -- ++(0,-0.5em) node[below] {\( \rrec \)};

        \draw[->, thick] (axis cs:36,0.73) -- (axis cs:36,0.28) node[midway,right,xshift=0.5em,align=left] {adversary removed};

        \draw[|-|] (axis cs:0,-0.3) -- node[midway,below] {live} (axis cs:23,-0.3);
        \draw[|->] (axis cs:35,-0.3) -- node[midway,below] {live} (axis cs:50,-0.3);
        \draw[|->] (axis cs:0,-0.5) -- node[midway,below] {safe} (axis cs:50,-0.5);
        
        \end{axis}
    \end{tikzpicture}
    \caption{The timeline of recovery from adversarial majority. Safety is required at all times and liveness must recover once the adversary is removed from the protocol by an external process and honest majority is restored.
    }
    \label{fig:model-timeline}
\end{figure}

\begin{figure*}
    \centering
    \subfloat[Freezing protocol\label{fig:better-safe-freezing}]{%
        \includegraphics[width=0.35 \textwidth,keepaspectratio]{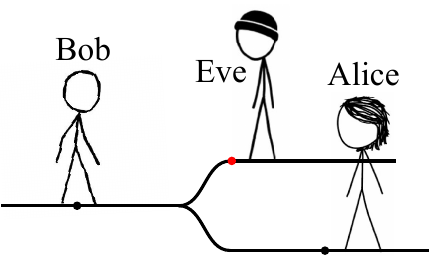}%
    }%
    \hspace{1cm}%
    \subfloat[Recoverable protocol\label{fig:better-safe-recovery}]{%
        \includegraphics[width=0.55 \textwidth,keepaspectratio]{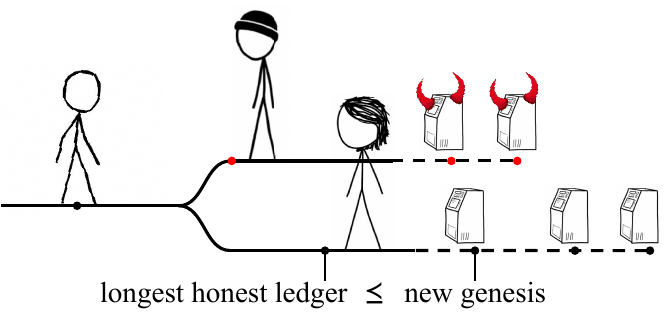}%
    }
    \caption{(a) By freezing the ledger, honest parties report ledgers that are prefixes of one another. However, Eve may confuse Bob, who holds a shorter ledger, as to what is its valid honest extension, \ie, the one confirmed by Alice. (b) A recoverable protocol makes a distinction between \emph{confirmed} ledgers (solid lines) and \emph{bookmarked} ledgers (dashed lines). The longest honestly confirmed ledger is a prefix of the shortest bookmarked ledger, and all of them are consistent with each other. The new genesis block can be computed by a majority vote between the bookmarked ledgers held by the validators (illustrated as machines).}
    \label{fig:better-safe-scenarios}
\end{figure*}

Even though we have achieved safety up to a $99\%$ adversary, the money is useless if it can never be spent again. We want liveness to eventually recover. 
To achieve this, the community who met at the bar uses evidence of misbehavior or other means to exclude the adversary from the protocol so that honest majority is restored (\cref{fig:model-timeline}).
Still, a challenge remains:
Regrettably, Alice and Bob may each freeze their ledger at a different length (\cref{fig:better-safe-freezing}). To unfreeze with human intervention, the community must make a choice of \emph{which ledger to extend}. If they extend the shorter one confirmed by Bob, then Alice, who confirmed a longer ledger, will be unhappy, as she will lose money. They must therefore extend the longer one. However, even though Alice knows in her heart that she's being truthful, the other honest parties may see conflicting evidence. In particular, Eve may claim that \emph{her} view of the ledger is the correct one, and honest parties who are behind cannot tell the difference (\cref{fig:better-safe-freezing}). The situation is frustrating: There \emph{is} a right choice to be made (the longest honest ledger), but this choice is unknowable, even if we try to resolve the conflict socially (see \cref{sec:protocol-freezing-no-recovery}).

It was previously believed~\cite{vitalik-blog} that such deadlocks are insurmountable and that an arbitrary choice must be made during those perilous circumstances. In this work, we challenge this folklore belief by introducing a new \emph{gadget} which transforms existing protocols to allow for recovery after honest majority heals. The resulting protocol is safe and live under honest majority. When the adversary gains majority (at the moment $\rmaj$, see timeline in \cref{fig:model-timeline}) and attempts to perform an infraction, our protocol freezes to preserve safety. When the community restores honest majority (for example, by slashing or through a social mechanism), our protocol recovers in a way that preserves safety. At the moment of recovery after honest majority heals ($\rrec$ in \cref{fig:model-timeline}), which must be truthfully announced as external advice, the protocol computes a \emph{new genesis} block evaluated based on past evidence. The new genesis block is guaranteed to extend the longest ledger confirmed by any honest party, even if the adversary presents confusing evidence. Beyond healing the population and announcing the moment of recovery, the computation of the new genesis block is automatic and internal to the protocol.

Crucial to this result is our network model. We make a distinction between clients and validators:
\emph{Validators} form a set of known online participants taking actions, such as voting, in the protocol. \emph{Clients} are unknown and free to join and leave the network at any time (sleepy clients)\footnote{Joining and leaving is similar to the sleepy model~\cite{sleepy}, but with the marked distinction between \emph{sleepy clients} and validators.}. 
Contrary to previous work, which only allowed validators to gossip messages to all parties, we also allow clients to do so.
In other words, any message received by a client will soon be received by all online parties. 
This is a more realistic depiction of currently deployed gossip networks comprised of both clients and validators.
The client gossip allows us to bypass previous impossibility results (see Related Work).

\noindent
\textbf{Our contributions.}
In summary, we make the following contributions:
\begin{enumerate}
  \item \emph{A freezing, always-safe protocol}.
  Our first, extremely simple, protocol is one that is always \emph{safe}, but \emph{live} only up to a $50\%$ adversary. It involves waiting before confirming a ledger, gossiping evidence of conflict, and freezing on observing evidence of conflict. Despite its straightforward design, it achieves something never seen in the literature before: an always-safe protocol which supports sleepy clients. In the analysis, we note two modeling intricacies:
  (a) \emph{safety} and \emph{liveness} resilience bounds do not need to be conflated; and
  (b) in real gossip networks, clients, too, can gossip messages.

  \item \emph{An always-safe recoverable protocol.} Our second protocol achieves safety always and liveness before adversarial majority, and after the population has healed from adversarial majority, recovers liveness in a safe manner (as shown in \cref{fig:model-timeline}). Remarkably, the protocol can support sleepy clients who inherit similar liveness and safety guarantees whenever they are awake. A new modeling aspect is the introduction of $\rrec$, a moment of recovery, in which the protocol receives trusted external advice that the population has healed, after which recovery can commence.
\end{enumerate}

\begin{figure}[tb]
    \centering
    \begin{tikzpicture}[node distance=2cm, auto,>=latex']

        \node[draw, rectangle, minimum size=1cm,inner sep=10pt] (box) {$\Pi$};

        \draw[<-] (box.west) -- ++(-1cm,0) node[midway, above] {$\txs$};

        \node[right=2cm of box, draw, rectangle, minimum size=0.5cm, inner sep=5pt] (circleDelta) {wait $\Delta$};
    
        \draw[->] (box.east) -- (circleDelta.west) node[midway, above, align=center] {ledger $\LOG$};

        \coordinate (point1) at ($(box.east)!0.3!(circleDelta.west)$);
        \coordinate (point2) at ($(box.east)!0.7!(circleDelta.west)$);
        \coordinate (midpoint) at ($(box.east)!0.5!(circleDelta.west)$);

        \node[below=2cm of midpoint, draw, cloud, cloud puffs=10, cloud puff arc=120, aspect=2, inner ysep=0] (cloud) {Network};
        \draw[<-] (point1) -- ($(point1 |- cloud.north) + (0,-0.1cm)$);
        \draw[->] (point2) -- ($(point2 |- cloud.north) + (0,-0.1cm)$) node[right, yshift=1em] {gossip};

        \node[right=0.75cm of circleDelta, regular polygon, regular polygon sides=3, shape border rotate=-90, draw, minimum size=0.5cm, inner sep=1pt] (triangle) {$\consistent$};

        \draw[->] (circleDelta.east) -- (triangle.west);

        \coordinate (midPoint2) at ($(circleDelta.west) + (-0.25cm, 0)$);
        \coordinate (triMidTop) at (triangle.north);
        \draw[->] (midPoint2) -- +(0,1cm) -| (triMidTop);
    
         \draw[->] (triangle.east) -| +(0.25cm,-0.4cm) node[yshift=-0.4cm, align=center] (logack) {$\LOGacki{}{}$};
         
         \node[below=0cm of logack, align=center, inner sep=0] (logacktext) {(bookmarked)};

        \node[right=0.75cm of triangle, draw, rectangle, minimum size=0.5cm, inner sep=5pt] (circleDelta2) {wait $2\Delta$};
    
        \draw[->] (triangle.east) -- (circleDelta2.west);
        \node[right=0.75cm of circleDelta2, regular polygon, regular polygon sides=3, shape border rotate=-90, draw, minimum size=0.5cm, inner sep=1pt] (triangle2) {$\consistent$};

        \draw[->] (circleDelta2.east) -- (triangle2.west);

        \coordinate (triMidTop2) at (triangle2.north);
        \draw[->] (midPoint2) -- +(0,1cm) -| (triMidTop2);
    
        \draw[->] (triangle2.east) -- +(1cm,0);
        \node[right=0.2cm of triangle2, yshift=-0.8cm, align=left] (logfinal) {$\LOGfinali{}{}$};
        \node[below=0cm of logfinal, align=left, inner sep=0, anchor=north west, xshift=-1em] (logfinaltext) {(confirmed)};

        \node[] (succeq) at ($(logack)!0.5!(logfinal)$) {$\succeq$};
        \draw[densely dotted, yshift=-0.1cm] (logack.east) -- (succeq.west);
        \draw[densely dotted, yshift=-0.1cm] (succeq.east) -- (logfinal.west);
    
        \node[draw, dashed, fit={(box.west) (triangle2.east) ($(midpoint) - (0,1.5cm)$) ($(midpoint) + (0,1.5cm)$)}, inner sep=0pt] (boundingBox) {};
    
        \node[below=0.1cm of boundingBox.south east, anchor=north east] {$\PI$};
    
    \end{tikzpicture}
    \caption[]{%
    \newcommand{\smallTriangle}{
        \tikz[baseline=-0.3em]{
            \node[regular polygon, regular polygon sides=3, shape border rotate=-90, draw, inner sep=0] {$\scriptscriptstyle\consistent$};
        }
    }
    A simplified block diagram of the recovery gadget. The internal protocol $\Pi$ is any protocol that is safe and live under honest majority.
    Ledgers output by $\Pi$ or obtained from the network are gossiped to the rest of the network.
    The component \smallTriangle remembers the set of all ledgers it ever received at the input port on its top.
    On receiving a ledger $\LOG$ at the input port on its left, this block outputs
    $\LOG$ if no conflicting ledgers were ever received.
    On waiting and checking for conflicts once, a ledger is considered \emph{bookmarked}.
    On waiting longer and checking conflicts again, it is considered \emph{confirmed}.
    Clearly, the bookmarked ledger extends the confirmed ledger.
    }
    \label{fig:recovery-protocol}
\end{figure}
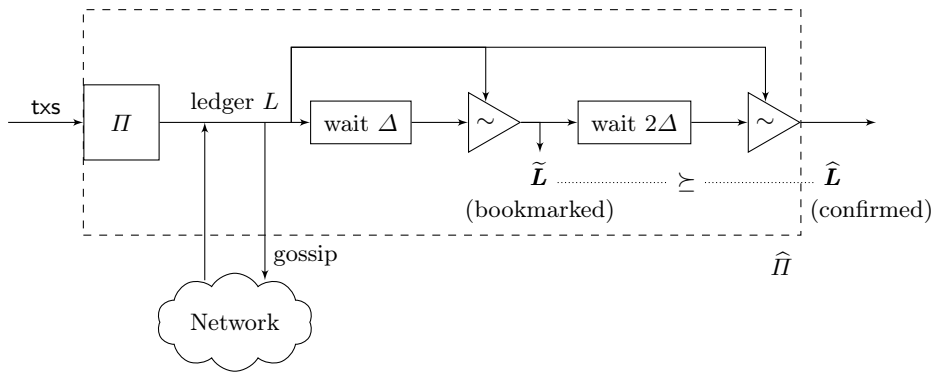

\noindent
\textbf{Construction overview.}
Our recoverable protocol constitutes a \emph{recovery gadget} built in a black-box fashion on top of any existing distributed ledger protocol (\cref{fig:recovery-protocol}) that provides certificates of confirmation and is secure under honest majority (such as Sync-Hotstuff~\cite{synchotstuff} and Sync-Streamlet~\cite{streamlet}). The \emph{internal} protocol $\Pi$ outputs ledgers that are consumed by the gadget. The gadget \emph{delays} these ledgers twice before reporting them to the external user: Once before they are considered \emph{bookmarked}; and, secondly, before they are considered \emph{confirmed}. Transactions are only considered confirmed when they are part of the confirmed ledger. Any conflicting ledgers that appear in the internal protocol at any point in time are used as a signal to freeze the protocol, and this evidence of infraction is gossiped to the rest of the network to help everyone safely freeze shortly thereafter too. This simple freezing idea ensures all \emph{bookmarked} and \emph{confirmed} ledgers of different honest parties are mutually consistent, ensuring safety even under dishonest majority. The relationship between \emph{bookmarked} and \emph{confirmed} ledgers is that the longest confirmed ledger of clients (solid line in \cref{fig:better-safe-recovery}) is behind the shortest bookmarked ledger of validators (dashed line in \cref{fig:better-safe-recovery}). When the moment for recovery arrives, and the population of validators (shown as machines in \cref{fig:better-safe-recovery}) has been healed of dishonest majority ($\rrec$ in \cref{fig:model-timeline}), the validators begin a voting phase in which they sign and broadcast their bookmarked ledgers. Clients listen for votes containing each validator's bookmarked ledger and take the prefix of them that is vouched by a majority. This guarantees, due to honest majority, to lead to a ledger (``new genesis'', left-most honest machine in \cref{fig:better-safe-recovery}) that contains the longest previously confirmed ledger. Subsequently, the gadget reboots the internal consensus protocol using the ledger discovered in this manner as a new starting point, safely recovering liveness.

\noindent
\textbf{Paper structure.} Our simple ``freezing'' idea is presented in \cref{sec:protocol-warmup} where we show that, with \emph{client gossip}, it achieves safety against up to a $99\%$ adversary.
We give a detailed description of the scenario of \cref{fig:better-safe-recovery} in \cref{sec:protocol-freezing-no-recovery} which motivates the steps required to
design our full recoverable protocol in \cref{sec:protocol-recovery}. We prove our protocol achieves the desired properties in \cref{sec:analysis}.
In \cref{sec:discussion}, we discuss extensions to proof-of-stake and partial synchrony.

\noindent
\textbf{Related work.}
The question of achieving resilience against majority adversaries is not new. Lamport \emph{et al.}~\cite{byzantine-generals} in 1982, and Dolev and Strong~\cite{dolev-strong} in 1983 already proposed a broadcast protocol achieving both safety and liveness with resilience up to $100\%$. Repeated executions make a consensus protocol~\cite{vitalik-dolev-strong,bcube,shi-adv-maj,flint-adv-maj}. Unfortunately, their protocol requires clients to be always online. 
It is well-known that with clients who may join and leave, it is impossible to achieve \emph{both} safety and liveness with more than $50\%$ resilience~\cite{schneider-survey,sleepy,shi-rethinking}.
Once safety and liveness are decoupled, \cite{aav1,mtbft} prove that safety resilience of $99\%$ would drive liveness resilience to $0\%$.
But the results of~\cite{aav1,mtbft} critically rely on the fact that clients cannot gossip messages they receive, an unrealistic view of real gossip networks.
By utilizing client gossip, our protocol achieves $99\%$ safety resilience and $50\%$ liveness resilience.
In our model of client gossip, although a client waking up receives messages sent while it was asleep, it cannot discern the round at which they were sent.
This means our model is weaker than a model with always-online clients.
All the above results, and our work, are for a synchronous network.
In a partially synchronous network, client gossip does not help because gossiped messages may not be delivered in time.
However, we discuss in \cref{sec:discussion} how our gadget preserves $33\%$ security under partial synchrony while recovering from a $99\%$ adversary under synchrony.
The different network models and the safety and liveness resilience achievable and impossible in each model are summarized in a table in \cref{sec:appendix-related}.
The technique of waiting and confirming on absence of conflict is also used by \emph{validators} in some protocols~\cite{synchotstuff,streamlet} with $50\%$ resilience, but client gossip is required to push safety to $99\%$, and even further changes are needed to allow recovery.

Situations of temporary adversarial majority and healing have been explored in both proof-of-work~\cite{zeta-pow-healing} and proof-of-stake~\cite{aggelos-pos-healing}. In these treatments, safety and liveness are both temporarily lost and recovered later, which can, regrettably, cause a loss of funds.
Some works~\cite{zlb,accountable-reconfigure} use accountability~\cite{fault-detection,bftforensics,aa,peerreview,polygraph} to automatically detect some (but not all) kinds of adversaries and remove them from the set of participants. These works, too, concede safety under adversarial majority.
In contrast, we assume that the adversary is removed, automatically or manually, and propose
the first
protocol that is always safe and recovers liveness.
\section{Model}
\label{sec:model}

\noindent
\textbf{Time.} Time proceeds in discrete \emph{rounds} indexed $r = 0, 1, 2, \ldots$. All parties have synchronized local clocks.

\noindent
\textbf{Parties.} There are two kinds of parties: \emph{validators} and \emph{clients}.
At round $0$, a set $\valset$ of validators awaken and remain awake until they are killed by the environment, whereas clients may awaken at any round, stay awake for an arbitrary number of rounds, and then sleep forever\footnote{A client that sleeps for a while and awakens again later is treated as if the client sleeps forever and a new client is awakened in its place.}. We denote $\rboot_k$ the round in which client $k$ awakens (for validators, $\rboot = 0$). 
The number of validators $n = |\valset|$ is known to everyone, whereas
the number of clients is unknown.

\noindent
\textbf{PKI.} Each validator has a public and private key. The public keys of all validators are known by all parties. All messages sent by a validator are signed using the validator's private key.

\noindent
\textbf{Distributed ledger protocol.} Validators and clients run a distributed ledger protocol $\Pi$.
The roles of a validator and client are different.
At each round, a validator reads inputs (called \emph{transactions}) from the environment and collaborates with other validators with the aim of placing these transactions into a total order.
A client does not receive transactions from the environment, but interacts with the validators and other clients, and at every round, 
confirms (outputs)
a sequence of transactions called the \emph{ledger}.
At round $0$, the protocol is initialized with the validator set $\valset$ and a genesis ledger (which may or may not be empty).
All ledgers output by the protocol must extend the genesis ledger.

\noindent
\textbf{Network.} Validators and clients can send messages to each other. The network is synchronous, \ie, a message sent by an honest party at round $r$ is delivered to honest party $p$ latest by round $\max\{r,\rboot_p\} + \Delta$.
Note that this means when a client wakes up at round $\rboot_p>0$, then it soon receives all messages that were sent to the network before it awoke\footnote{In practice, this can be achieved by having online parties send all important past messages, including equivocations, to a newly-joining client.}, but the adversary may also inject some of its own messages (similar to the model in~\cite{sleepy}).
However, different validators or clients may receive a message at different rounds.

The following aspects of the model are new to this work:

\noindent
\textbf{Client gossip.}
In our network model described above,
every honest validator and client re-broadcasts every message it receives to the rest of the network. In deviation from previous literature, we even allow \emph{clients} to gossip messages they receive, and validators to possibly act on such messages. On the one hand, this is a more accurate depiction of reality in which communication is facilitated by a non-eclipsed gossip network comprised of both clients and validators.
On the other hand, this seemingly insignificant change in the network model enables us to circumvent impossibility results~\cite{aav1,mtbft}.

\noindent
\textbf{Time-varying adversarial corruption.} 
Since our work deals with recovering from adversarial majority, we model an adversary whose corruption level varies over time.
We denote by $f(r)$ the fraction of validators that are corrupted by the adversary at round $r$.
As seen in \cref{fig:model-timeline}, there are two important rounds chosen by the adversary, adaptively: $\rmaj$ (the \emph{adversarial majority} round) and $\rrec > \rmaj$ (the \emph{recovery round}).
Before $\rmaj$, the adversary maintains $f(r) < \frac{1}{2}$, while during $[\rmaj,\rrec)$, the adversary may corrupt $f(r) \geq \frac{1}{2}$.
Honest parties do not know $\rmaj$.

Only at round $\rrec$, the environment is allowed to kill validators to ensure that $f(\rrec) < \frac{1}{2}$, which we describe in the paragraph below.
After $\rrec$, the adversary may corrupt more validators, but 
it is assumed that $f(r) < \frac{1}{2}$ for $r \geq \rrec$.
A validator or client, once corrupted, remains corrupted forever, so $f(r)$ is non-decreasing in $r$, except at round $\rrec$.
Similar to the eventual synchrony model~\cite{dls88} in which a single transition from asynchrony to synchrony (GST) is a proxy for alternating periods of asynchrony and synchrony, we use $\rmaj$ and $\rrec$ as a proxy for alternating period of honest majority, adversarial majority, and recovery.

Since, in our model, clients can influence the execution (through client gossip), we also allow the adversary to corrupt any number of clients.
The adversary has access to the internal state of all corrupted parties, including private keys.

\noindent
\textbf{Healing honest majority.} 
At round $\rrec$,
the environment 
must kill a number of validators such that less than $\frac{1}{2}$ of the \emph{remaining} validators are corrupted. This causes the number of validators $n$ to be updated to a new number of validators $n' \leq n$. 
In practice, this cleanup may take place by internal \emph{slashing}~\cite{vitalik-blog} by the protocol (c.f., \emph{tombstoning}~\cite{cosmos-tombstone}) or by a social consensus process in which adversarial validators are removed by the community.
This is akin to classic models of external reconfiguration~\cite{spiegelman-reconfiguration}. 

At round $\rrec$, the environment sends a special message $\langle \text{recover}, \valsetnew \rangle$ to all parties announcing that recovery must now commence. This message can only be sent at round $\rrec$ and includes the public keys of the set $\valsetnew$ of the $n'$ surviving validators.
This allows the honest parties to update their PKI.
This special message is also sent to all clients that wake up after $\rrec$.
An external update of the PKI is, in fact, a software update that can be realized through a ``hard fork''~\cite{vitalik_sbc22}.
Thus, we follow previous literature~\cite{updatable}, which modeled coordination of software updates by a special message sent by the environment announcing the code of the updated protocol and the moment at which the update takes effect.
In practice, agreement on such an announcement can be achieved by a community vote external to the protocol~\cite{zhang-ndss}.

\noindent
\textbf{Notation.} We use $A \preceq B$ to denote that sequence $A$ is a (not necessarily strict) prefix of the sequence $B$. We use $A \consistent B$ as a shorthand for $A \preceq B \,\lor B\, \preceq A$.
The ledger confirmed by party $p$ at round $r$ is denoted $\LOGfinali{p}{r}$.

\noindent
\textbf{Security.} We define the following properties for the protocol:

\begin{definition}[Safety]
    \label{def:safety}
    A distributed ledger protocol $\Pi$ is \emph{safe} 
    in a set of rounds $I$
    if
    for all rounds $r,s \in I$ and all honest clients $p,q$ awake at rounds $r,s$ respectively, $\LOGfinali{p}{r} \consistent \LOGfinali{q}{s}$.
\end{definition}

\begin{definition}[Liveness]
    \label{def:liveness}
    A distributed ledger protocol $\Pi$ is \emph{live} with latency $u$
    in a set of rounds $I$
    if
    for all rounds $r \in I$,
    if a transaction $\tx$ was received by all honest validators before round $r - u$, 
    then for all honest clients $p$ with $\rboot_p < r - u$,
    $\tx \in \LOGfinali{p}{r}$.
\end{definition}

In this work, we develop a protocol that safely recovers from adversarial majority as defined below.

\begin{definition}
\label{def:rec-liveness}
    A distributed ledger protocol $\Pi$ is said to \emph{safely recover from adversarial majority}
    over an execution of length $\Rhrzn$
    if 
    for some finite $u,\urec$,
    $\Pi$ is safe in rounds $[0,\Rhrzn]$ and live with latency $u$ in rounds $[0,\rmaj) \cup (\rrec+\urec,\Rhrzn]$.
\end{definition}
\section{Protocol}
\label{sec:protocol}

In this section, we describe the recovery gadget protocol.
As a warmup, we start with the freezing gadget
which preserves safety under adversarial majority, but does not permit recovery after honest majority heals.
The full recovery gadget builds on this idea and uses the same key components as the freezing gadget.

These gadgets are pieces of code that are meant to be run
as an add-on to
a distributed ledger protocol $\Pi$
that provides safety and liveness under honest majority.
We will call $\Pi$ the \emph{internal protocol}.
The gadget then specifies actions that clients and validators perform using the output of $\Pi$ (see \cref{fig:freezing-protocol}).
The gadget, when combined with $\Pi$, also forms a distributed ledger protocol (validators input transactions and clients confirm a ledger), that we call the \emph{freezing protocol} $\PI$ (\cref{fig:freezing-protocol}), that provides safety even under adversarial majority. In other words, our constructions are by computational reduction\footnote{Formally speaking, because protocols are executed by multiple parties within an execution, these reductions are more complicated. They can be modelled as a subprotocol in the form of an Interactive Turing Machine Instance in the UC setting.}.
We will denote the internal protocol's output ledger as $\LOGi{}{}$ and the freezing protocol's as $\LOGfinali{}{}$. 
Both the freezing gadget and recovery gadget apply to protocols that satisfy a property called \emph{certifiability}~\cite{aa,roughgarden}.
Several PBFT-style protocols such as HotStuff~\cite{hotstuff,hotstuff2}, Streamlet~\cite{streamlet}, Tendermint~\cite{tendermint}, Casper~\cite{casper}, and their synchronous variants such as Sync HotStuff~\cite{synchotstuff} and Sync-Streamlet~\cite[Sec.~4]{streamlet}
have this property.
In these protocols, clients confirm a ledger on seeing a certain number of blocks with quorum certificates, which forms a \emph{\transcript}.
Any other client, on seeing this \transcript, irrespective of other messages that it may have seen, considers this ledger to be confirmed.
Such protocols have a stronger notion of safety: a minority adversary cannot create \transcripts certifying two conflicting ledgers.
In our gadgets, clients use \transcripts to let each other know about conflicting ledgers that could potentially be confirmed so that they freeze instead of confirming.
In what follows, we consider certifiable protocols that are secure under synchrony and honest majority, \eg, Sync-HotStuff~\cite{synchotstuff} and Sync-Streamlet~\cite[Sec.~4]{streamlet}.\footnote{These protocols can be made certifiable by having validators broadcast a signature on their committed/finalized ledgers~\cite[Sec.~4.2]{mtbft}.}

\begin{definition}[{Certifiable protocol}]
\label{def:certifiable}
    A distributed ledger protocol $\Pi$
    accompanied by 
    a computable functionality $\transcribe$ (the \emph{witness producer})
    and
    a computable deterministic non-interactive function
    $\untranscribe$ (the \emph{witness consumer}),
    is called \emph{certifiable}.
    When a client $p$ invokes $\transcribe()$ at round $r$, it produces a \transcript $\transcript$ such that $\untranscribe(\transcript) = \LOGi{p}{r}$.
\end{definition}

\begin{definition}[Certifiable safety]
\label{def:certifiable-safety}
    A certifiable protocol $\Pi$ accompanied by $\transcribe$ and $\untranscribe$ is certifiably safe in a set of rounds $I$ if
    $\Pi$ is safe in $I$,
    and moreover,
    if at any round $r \in I$,
    the adversary outputs a \transcript $\transcript$ such that $\untranscribe(\transcript) = \LOG$,
    then for all clients $q$, for all rounds $s \in I$,
    $\LOG \consistent \LOGi{q}{s}$.
\end{definition}

In the following, we will consider any certifiable protocol that is certifiably safe and live under honest majority (\cref{def:internal-protocol-security}), in particular safe and live until $\rmaj$, as the internal protocol.

\subsection{Always Safe: The \emph{Freezing} Gadget}
\label{sec:protocol-warmup}

We begin by introducing our freezing gadget.
Whereas the construction is simple, the result of an always safe protocol is remarkable: Different honest parties can never reach conflicting conclusions.

\begin{algorithm}
\caption{Freezing Gadget (run by clients)}
\label{alg:freezing}
\begin{algorithmic}[1]
\footnotesize
\On{\Call{init}{$\valset,\genesis$}}
\label{loc:freezing-init}
    \State $P \gets \text{new } \Pi(\valset, \genesis)$ \Comment{instantiate a new $\Pi$ client} \label{loc:freezing-base-protocol}
    \State $\msgSet \gets \emptyset$ \Comment{set of valid ledgers seen so far}
    \State $\LOGfinali{}{} \gets \genesis$ \Comment{confirmed (output) ledger of the combined protocol $\PI$}
\EndOn

\On{$\transcript$ output by $P.\transcribe()$ once per round or $\transcript$ received from network} \label{loc:freezing-on-witness}
    \State $\LOG \gets \untranscribe(\transcript)$ \label{loc:freezing-verify}
    \State $\msgSet \gets \msgSet \cup \{\LOG\}$
    \State $\operatorname{gossip}(\transcript)$ \label{loc:freezing-gossip} %
    \State $\operatorname{wait}(\Delta)$ \Comment{meanwhile, continue processing other events} \label{loc:freezing-wait}
    \If{$\LOG \not\preceq \LOGfinali{}{} \land \forall \LOG' \in \msgSet \colon \LOG \consistent \LOG'$} \Comment{ledger has grown, no conflicting ledgers} \label{loc:freezing-check-conflict}
        \State $\LOGfinali{}{} \gets \LOG$ \label{loc:freezing-output}
    \EndIf
\EndOn
\end{algorithmic}
\end{algorithm}

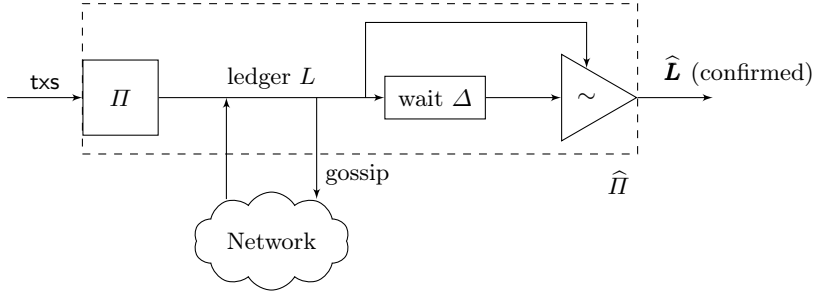
\begin{figure}[tb]%
    \centering%
    \begin{tikzpicture}[node distance=2cm, auto,>=latex']

    \node[draw, rectangle, minimum size=1cm,inner sep=10pt] (box) {$\Pi$};

    \draw[<-] (box.west) -- ++(-1cm,0) node[midway, above] {$\txs$};

    \node[right=3cm of box, draw, rectangle, minimum size=0.5cm, inner sep=5pt] (circleDelta) {wait $\Delta$};

    \draw[->] (box.east) -- (circleDelta.west) node[midway, above, align=center] {ledger $\LOG$};

    \coordinate (point1) at ($(box.east)!0.3!(circleDelta.west)$);
    \coordinate (point2) at ($(box.east)!0.7!(circleDelta.west)$);
    \coordinate (midpoint) at ($(box.east)!0.5!(circleDelta.west)$);

    \node[below=1.25cm of midpoint, draw, cloud, cloud puffs=10, cloud puff arc=120, aspect=2, inner ysep=0] (cloud) {Network};
    \draw[<-] (point1) -- ($(point1 |- cloud.north) + (0,-0.1cm)$);
    \draw[->] (point2) -- ($(point2 |- cloud.north) + (0,-0.1cm)$) node[right, yshift=1em] {gossip};

    \node[right=1cm of circleDelta, regular polygon, regular polygon sides=3, shape border rotate=-90, draw, minimum size=1cm] (triangle) {$\consistent$};

    Draw an arrow from circleDelta to the triangle
    \draw[->] (circleDelta.east) -- (triangle.west);

    \coordinate (midPoint2) at ($(circleDelta.west) + (-0.25cm, 0)$);
    \coordinate (triMidTop) at (triangle.north);
    \draw[->] (midPoint2) -- +(0,1cm) -| (triMidTop);

    \draw[->] (triangle.east) -- +(1cm,0) node[midway, above, anchor=south west, yshift=0.15cm, xshift=-0.15cm, inner sep=0] {$\LOGfinali{}{}$ (confirmed)};

    \node[draw, dashed, fit={(box.west) (triangle.east) ($(midpoint) - (0,0.75cm)$) ($(midpoint) + (0,1.25cm)$)}, inner sep=0pt] (boundingBox) {};

    \node[below=0.1cm of boundingBox.south east, anchor=north east] {$\PI$};

\end{tikzpicture}
    \caption[]{%
        \newcommand{\smallTriangle}{
            \tikz[baseline=-0.3em]{
                \node[regular polygon, regular polygon sides=3, shape border rotate=-90, draw, inner sep=0] {$\scriptscriptstyle\consistent$};
            }
        }
        The freezing gadget.
        The internal protocol $\Pi$ is any certifiable distributed ledger protocol.
        On seeing a ledger $\LOG$ (formally, seeing a \transcript $\transcript$ such that $\untranscribe(\transcript) = \LOG$) output by $\Pi$ or from the network, the client gossips $\LOG$ (formally, $\transcript$) to the rest of the network, and then waits for $\Delta$ rounds.
        The \emph{conflict resolution} component \smallTriangle remembers the set $\msgSet$ of all ledgers it ever received at the input port on its top.
        On receiving $\LOG$ at the input port on its left, this component outputs
        $\LOG$ if there were no conflicting ledgers in $\msgSet$
        (see \cref{alg:freezing}~\cref{loc:freezing-check-conflict,loc:freezing-output}).
        The ledger confirmed by the freezing protocol $\PI$ is denoted $\LOGfinali{}{}$.
    }%
    \label{fig:freezing-protocol}%
\end{figure}%

The freezing gadget is described in \cref{alg:freezing} and is illustrated as a block diagram in \cref{fig:freezing-protocol}.
This gadget, when applied to a certifiable distributed ledger protocol $\Pi$ (the internal protocol), produces a distributed ledger protocol $\PI$ (the freezing protocol).
The freezing gadget is only run by clients.
A client for the freezing protocol $\PI$ internally runs a client for the internal protocol $\Pi$ ($\Pi$ in \cref{fig:freezing-protocol}, \cref{alg:freezing}~\cref{loc:freezing-base-protocol}).
All parties are connected to the network (as modeled in \cref{sec:model}).

Each client uses the $\transcribe()$ functionality of the certifiable protocol to periodically output a \transcript $\transcript$.
It may also receive \transcripts in the form of messages sent by honest or adversarial parties in the network.
The client extracts a ledger $\LOG$ from the \transcript (\cref{alg:freezing}~\cref{loc:freezing-verify}),
adds $\LOG$ to its set of seen ledgers $\msgSet$ and gossips the \transcript (\cref{alg:freezing}~\cref{loc:freezing-gossip}).
The client then waits for $\Delta$ rounds, during which it continues to process other received \transcripts in the same manner (\cref{alg:freezing}~\cref{loc:freezing-wait}).
At the end of the wait, the client confirms the ledger $\LOG$ if it has seen no conflicting ledgers and if $\LOG$ is longer than its previously confirmed ledger. (\cref{alg:freezing}~\cref{loc:freezing-check-conflict,loc:freezing-output}).
Notice that when one client sees conflicting ledgers $\LOG, \LOG'$, after $\Delta$ rounds, all clients have added both $\LOG$ and $\LOG'$ to their respective $\msgSet$ sets, and will thereafter not confirm any new ledger, \ie, they \emph{freeze}.
However, crucially, they continue to gossip messages and \transcripts they see after they have frozen.

\begin{figure}[tb]
    \centering
    \begin{tikzpicture}

        \draw (0,0) -- (8,0) node[right] {round}; %
        \draw[->] (7.5,0) -- (8,0); %

        \coordinate (r) at (6,0);
        \coordinate (rdelta) at (2,0);
        \draw (rdelta) +(0,0.1) -- +(0,-0.1) node[below, anchor=base, yshift=-1em] {$r-\Delta$};
        \draw (r) +(0,0.1) -- +(0,-0.1) node[below, anchor=base, yshift=-1em] {$r$};

        \draw [Latex-,dashed] ([yshift=0.2cm]r.north) -- ([yshift=1.2cm]r.north) node [anchor=south west,align=left,xshift=-1em] {\DiagramStep{1} $p$ confirms $\LOG$};

        \draw [Latex-,dashed] ([yshift=0.2cm]rdelta.north) -- ([yshift=0.9cm]rdelta.north) node [anchor=south west,align=left,xshift=-1em] {\DiagramStep{2} $p$ sees $\LOG$,\\gossips $\LOG$};

        \draw [Latex-,dashed] ([yshift=0.2cm]r.north) |- ([yshift=0.7cm,xshift=0.7cm]r.north) node [anchor=west,align=left,xshift=0em] {\DiagramStep{3} all clients see $\LOG$,\\do not confirm $\LOG' \inconsistent \LOG$};
        
    \end{tikzpicture}
    \caption[]{Illustration for the freezing protocol's safety which is maintained during adversarial majority (\cref{lem:freezing-safety}).
    \DiagramStep{1}~Suppose that at round $r$, a client $p$ confirms a ledger $\LOG$.
    \DiagramStep{2}~The client must have seen $\LOG$ (formally, a \transcript $\transcript$) either from the internal protocol $\Pi$ or from the network
    latest by round $r - \Delta$, at which point it must have gossipped $\LOG$
    (formally, $\transcript$). 
    \DiagramStep{3}~Thus, by round $r$, all clients must have seen $\LOG$ 
    and thereafter will never confirm a ledger that conflicts with $\LOG$.
    }
    \label{fig:freezing-safety-proof}
\end{figure}
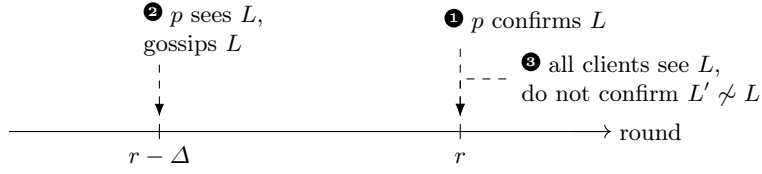

The freezing protocol provides safety and liveness until round $\rmaj$ and retains safety after $\rmaj$. We prove these properties below.
The freezing protocol's safety comes entirely from the freezing gadget.
See \cref{fig:freezing-safety-proof} for a visual summary of the proof.
Liveness during honest majority is a consequence of liveness and safety of the internal protocol.
Liveness of the internal protocol ensures that new transactions are included in its output.
Safety of the internal protocol ensures that honest clients do not see \transcripts for conflicting ledgers, hence they eventually confirm all ledgers output by $\Pi$.

\begin{lemma}[Safety]
\label{lem:freezing-safety}
For any set of rounds $I$, $\PI$ is safe in $I$.
\end{lemma}

\begin{proof}
See \cref{fig:freezing-safety-proof} for reference.
    Towards contradiction, let $r$ be the smallest round such that for some $s \geq r$, and some honest clients $p,q$, $\LOGfinali{p}{r} \inconsistent \LOGfinali{q}{s}$.
    For shorthand, let $\LOG = \LOGfinali{p}{r}$.
    Then, at round $r-\Delta$, client $p$ must have 
    seen a \transcript $\transcript$ such that $\untranscribe(\transcript) = \LOG$.
    Client $p$ also gossiped 
    $\transcript$
    at round $r-\Delta$, which means that before the end of round $r$, client $q$ must have 
    seen $\transcript$.
    Thus, client $q$ added $\LOG$
    to its set $\msgSet$ before the end of round $r$.
    However, since client $q$ confirmed $\LOGfinali{q}{s} \inconsistent \LOG$ at round $s \geq r$, this is a contradiction to the freezing (\cref{alg:freezing}~\cref{loc:freezing-check-conflict}).
\end{proof}

\begin{lemma}[Liveness]
\label{lem:freezing-liveness}
If $\Pi$ is certifiably safe and live with latency $u_{\Pi}$ in rounds $[0,\rmaj)$, then $\PI$ is live with latency $u_{\Pi} + \Delta$ in rounds $[0,\rmaj)$.
\end{lemma}

\begin{proof}
    Let $u = u_{\Pi} + \Delta$.
    Let $r < \rmaj$ be any arbitrary round.
    Suppose that a transaction $\tx$ is received by all honest validators before round $r - u$.
    Consider an honest client $p$ that wakes up before $r - u$, \ie, $\rboot_p < r - u$.
    Due to liveness of $\Pi$, at round $s = r - u + u_{\Pi}$, 
    $\tx \in \LOGi{p}{s}$ (the ledger output by the internal protocol $\Pi$).
    At round $s$, client $p$ runs $\transcript \gets \transcribe()$ and adds $L = \untranscribe(\transcript)$ to its set $\msgSet$(\cref{alg:freezing}~\cref{loc:freezing-on-witness,loc:freezing-verify}).
    Recall from \cref{def:certifiable} that $\LOG = \LOGi{p}{s}$.
    Due to certifiable safety, 
    the set $\msgSet$
    of client $p$, 
    at all rounds before $\rmaj$, contains only ledgers that are consistent with $\LOG$.
    Therefore, at round $s + \Delta = r$, $\LOGfinali{p}{r} \succeq \LOG \ni \tx$ (due to \cref{alg:freezing}~\cref{loc:freezing-check-conflict,loc:freezing-output}).
\end{proof}

\subsection{Towards Recovery}
\label{sec:protocol-freezing-no-recovery}

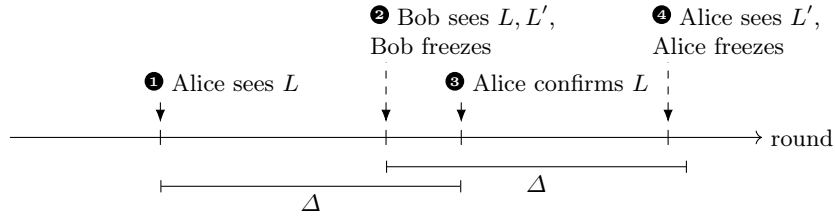
\begin{figure}[tb]
    \centering
    \begin{tikzpicture}

        \draw[->] (0,0) -- (10,0) node[right] {round}; %
        \coordinate (bsee) at (5,0);
        \coordinate (asee1) at (2,0);
        \coordinate (asee2) at (8.75,0);
        \coordinate (aconf) at (6,0);
        \coordinate (bconf) at (9,0);

        \draw (asee1) +(0,0.1) -- +(0,-0.1) node[below, anchor=base, yshift=-1em] {};
        \draw (bsee) +(0,0.1) -- +(0,-0.1) node[below, anchor=base, yshift=-1em] {};
        \draw (asee2) +(0,0.1) -- +(0,-0.1) node[below, anchor=base, yshift=-1em] {};
        \draw (aconf) +(0,0.1) -- +(0,-0.1) node[below, anchor=base, yshift=-1em] {};

        \draw [Latex-,dashed] ([yshift=0.2cm]asee1.north) -- ([yshift=0.5cm]asee1.north) node [anchor=south west,align=left,xshift=-1em] {\DiagramStep{1} Alice sees $\LOG$};

        \draw [Latex-,dashed] ([yshift=0.2cm]bsee.north) -- ([yshift=1.0cm]bsee.north) node [anchor=south west,align=left,xshift=-1em] {\DiagramStep{2} Bob sees $\LOG,\LOG'$,\\Bob freezes};

        \draw [Latex-,dashed] ([yshift=0.2cm]aconf.north) -- ([yshift=0.5cm]aconf.north) node [anchor=south west,align=left,xshift=-1em] {\DiagramStep{3} Alice confirms $\LOG$};
        
        \draw [Latex-,dashed] ([yshift=0.2cm]asee2.north) -- ([yshift=1.0cm]asee2.north) node [anchor=south west,align=left,xshift=-1em] {\DiagramStep{4} Alice sees $\LOG'$,\\Alice freezes};
        
        \draw [|-|] ([yshift=-2em]asee1) -- ([yshift=-2em]aconf) node [pos=0.5,below] {$\Delta$};
        \draw [|-|] ([yshift=-1.2em]bsee) -- ([yshift=-1.2em]bconf) node [pos=0.5,below] {$\Delta$};

    \end{tikzpicture}
    \caption[]{%
        Timeline of a scenario in which the freezing protocol results in two clients (Alice and Bob) freezing and the last ledger they confirmed are of different lengths.
        \DiagramStep{1}~First, client Alice sees a \transcript for a ledger $\LOG$ (in short, we say it ``sees $\LOG$'').
        \DiagramStep{2}~Client Bob also sees $L$ within $\Delta$ rounds.
        In addition, Bob sees another ledger $\LOG'$ which is inconsistent with $\LOG$. As a result, Bob does not confirm $\LOG$.
        \DiagramStep{3}~Meanwhile, Alice confirms $\LOG$ on waiting $\Delta$ rounds after seeing $\LOG$.
        \DiagramStep{4}~Eventually, Alice also sees $\LOG'$ (due to client gossip) and Alice freezes.
        The result is that Alice froze after confirming $\LOG$ but Bob froze before confirming $\LOG$, so they have ledgers that are consistent but of different lengths (as in \cref{fig:better-safe-freezing}).
    }
    \label{fig:freezing-no-recovery}
\end{figure}

Given that all clients have frozen, the protocol must recover liveness after $\rrec$.
One way to do this
is for the new validator set after $\rrec$ to restart the protocol
from a ``new genesis''.
However, to maintain safety for all clients, it is required that the new genesis contains a ledger that extends the previously confirmed ledgers of all clients.
In the scenario described in \cref{fig:better-safe-freezing}, although clients froze to maintain safety, the protocol cannot recover because honest validators are unable to decide on such a ledger.
A timeline of events leading up to this scenario is shown in \cref{fig:freezing-no-recovery}.
Due to network delay, different clients see messages at different rounds and in different orders.
As a result, while one client, Alice, confirmed a longer ledger before freezing, another client, Bob, may have confirmed only a prefix of Alice's ledger before freezing (as in \cref{fig:better-safe-freezing}).

Knowing that the freezing protocol provides safety against up to $100\%$ adversary, Alice knows that no other honest client would have confirmed any ledger that conflicts with the one she confirmed.
However, Bob (whose ledger lags behind) sees two conflicting ledgers, $\LOG$ and $\LOG'$, and does not know which of them may have been confirmed by honest clients.
Bob could attempt to find out by asking all other clients which ledger they confirmed.
However, there might even be \emph{adversarial clients} who claim that they confirmed $\LOG'$ (Eve in \cref{fig:better-safe-freezing}), while honest clients like Alice confirmed $\LOG$.
Bob, based on his view, cannot tell which client, Eve or Alice, is lying.
Since there may be an arbitrary number of adversarial clients, Bob cannot use a majority vote to decide which ledger was confirmed by honest clients.
One might consider running a majority vote on the set of validators, of whom we know that a majority are honest after round $\rrec$.
However, it might be the case that all validators have a view similar Bob's, \ie, they cannot tell which ledger was confirmed by honest clients.
Thus, even if validators were to run an interactive protocol to reach a consensus on which ledger was confirmed by honest clients, they would not be able to do so.
Not knowing which of the two ledgers to extend with new transactions, validators are unable to recover the protocol.

\subsection{The \emph{Recovery} Gadget}
\label{sec:protocol-recovery}

\begin{algorithm}[tb]
\caption{Recovery Gadget}
\label{alg:protocol}
\begin{algorithmic}[1]
\footnotesize
\LineComment{Code for validators}
\LineComment{Part 1: Same as freezing gadget (\cref{alg:freezing}~\crefrange{loc:freezing-init}{loc:freezing-output}), except replace $\LOGfinali{}{}$ by $\LOGacki{}{}$} \label{loc:recovery-val-code-copy}
\LineComment{Part 2 (Recovery):}
\On{\valcode{receiving $\langle \text{recover}, \valsetnew \rangle$ from the environment}} \Comment{\valcode{received at round $\rrec$}}
    \State \valcode{$\msgSetRec \gets \emptyset$} \Comment{\valcode{set of messages delivered during recovery}}
    \State \valcode{$\Call{broadcast}{\langle \text{bookmark}, \LOGacki{}{} \rangle}$} \label{loc:recovery-broadcast-bookmark}
    \State \valcode{$\operatorname{wait}(\bclatency)$} \Comment{\valcode{wait to deliver bookmarks broadcast by validators in $\valsetnew$, see \cref{loc:recovery-deliver-bookmark}}}
    \State \valcode{$\msgSetRec \gets \{\LOG \in \msgSetRec \colon |\{ \LOG' \in \msgSetRec \colon \LOG' \succeq L\}| > |\valsetnew|/2\}$ } \label{loc:recovery-new-genesis1}
    \State \valcode{$\LOGrec \gets \arg\max_{L \in \msgSetRec} |\LOG|$} \Comment{\valcode{longest prefix of ledgers bookmarked by a majority}} \label{loc:recovery-new-genesis}
    \State \valcode{$\operatorname{gossip}(\langle \text{genesis}, \LOGrec \rangle)$} \Comment{\valcode{declare the new genesis for clients joining after $\rrec$}} \label{loc:recovery-gossip-genesis}
    \State \valcode{$\Call{init}{\valsetnew, \LOGrec}$} \Comment{\valcode{restart with new validator set and new genesis}} \label{loc:recovery-val-restart}
\EndOn
\On{\valcode{$\Call{deliver}{\langle \text{bookmark}, \LOG \rangle}$ broadcast by $V \in \valsetnew$}}\Comment{At most one value per $V$} \label{loc:recovery-deliver-bookmark}
    \State \valcode{$\msgSetRec \gets \msgSetRec \cup \{\LOG\}$}
\EndOn
\medskip
\LineComment{Code for clients}
\LineComment{Part 1: Same as freezing gadget (\cref{alg:freezing}~\crefrange{loc:freezing-init}{loc:freezing-output}), except replace $\Delta$ by $3\Delta$} \label{loc:recovery-client-code-copy}
\LineComment{Part 2 (Recovery):}
\On{\clientcode{receiving $\langle \text{recover}, \valsetnew \rangle$ from the environment}} \Comment{\clientcode{received at $\max\{\rrec,\rboot\}$}}
    \State \clientcode{$\textsf{recovering} \gets \textsf{true}$}
    \State \clientcode{$\operatorname{lock}(\LOGfinali{}{})$} \Comment{forbid future updates to $\LOGfinali{}{}$} \label{loc:recovery-client-freeze}
    \State \clientcode{$\mathcal{S}_\text{gen} = \{~\}$}
\EndOn
\On{\clientcode{receiving $\langle \text{genesis}, \LOG \rangle$ from $V \in \valsetnew$ and $\textsf{recovering} = \textsf{true}$}}
    \State \clientcode{$\mathcal{S}_\text{gen}[\LOG] =  \mathcal{S}_\text{gen}[\LOG] + 1$}
    \If{$|\mathcal{S}_\text{gen}[\LOG]| > |\valsetnew|/2$}
        \State \clientcode{$\textsf{recovering} \gets \textsf{false}$}
        \State \clientcode{$\LOGrec \gets \LOG$} \label{loc:recovery-client-genesis}
        \State \clientcode{$\operatorname{unlock}(\LOGfinali{}{})$} \Comment{permit updates to $\LOGfinali{}{}$} \label{loc:recovery-client-unfreeze}
        \State \clientcode{$\Call{init}{\valsetnew,\LOGrec}$} \Comment{\clientcode{restart with new validator set and new genesis}} \label{loc:recovery-client-restart}
    \EndIf
\EndOn
\end{algorithmic}
\end{algorithm}

The recovery gadget (\cref{alg:protocol}) has two parts.
Part 1 is a protocol that is run at all times and is very similar to the freezing gadget, except that both clients and validators run the recovery gadget in different ways.
Part 1 is designed to provide a property in addition to safety during adversarial majority that allows honest validators to know which ledgers honest clients could have potentially confirmed.
Part 2 is a protocol that is performed only during recovery (which is announced by the environment at round $\rrec$), in which validators collaborate to decide on a new genesis from which to restart the protocol, and clients restart their protocol based on the validators' decision.

In Part 1 (\cref{fig:recovery-protocol}),
the code that validators run is identical to that of the freezing gadget, except that validators don't consider the output ledger as confirmed (recall, only clients confirm ledgers).
Instead, they consider a ledger as `bookmarked' (denoted $\LOGacki{}{}$) when it is output by the freezing gadget (\cref{fig:recovery-protocol}, \cref{alg:protocol}~\cref{loc:recovery-val-code-copy}).
The bookmarked ledgers are internal to the recovery gadget and their role is to assist validators during recovery.
The bookmarked ledgers of different validators are consistent before $\rrec$ (due to the freezing gadget's safety).
But the bookmarked ledgers after recovery may not remain consistent with those before recovery.
However, this doesn't matter since bookmarked ledgers are not confirmed by any party.

The clients' code during Part 1 is also identical to the freezing gadget, except that the client waits $3\Delta$ before confirming a ledger (\cref{alg:protocol}~\cref{loc:recovery-client-code-copy}), longer than the $\Delta$ that validators wait before bookmarking.
This is done in order to provide a property called \emph{follow-the-leader}.
Intuitively, by waiting longer to confirm, a client ensures that when it confirms a ledger, all honest validators have already bookmarked it.
Thus, at the time of recovery, every honest validator knows that it is safe to restart the protocol from the bookmarked ledger in its view because it is an extension of all clients' confirmed ledgers.
Thus, the longest common prefix of all honest validators' bookmarks is a safe new genesis (see \cref{fig:better-safe-recovery}).
Recall that $\LOGfinali{}{}$ denotes a client's confirmed ledger, and $\LOGacki{}{}$ a validator's bookmarked ledger.

\begin{definition}[Follow-the-leader]
\label{def:follow-the-leader}
    A distributed ledger protocol $\Pi$ has the follow-the-leader property if
    for all rounds $r$, clients $p$, and validators $v$, $\LOGfinali{p}{r} \preceq \LOGacki{v}{r}$.
\end{definition}

A validator begins running Part 2 of the recovery gadget when it receives a ``recover'' message from the environment at round $\rrec$, specifying the new validator set $\valsetnew$.
At this point, each validator broadcasts their own bookmarked ledger (\cref{alg:protocol}~\cref{loc:recovery-broadcast-bookmark}).
Upon receiving these bookmarks, 
each validator decides the new genesis $\LOGrec$ as the longest prefix of ledgers bookmarked by a majority of the new validator set (\cref{loc:recovery-new-genesis1,loc:recovery-new-genesis}).
Due to honest majority among the new validator set, $\LOGrec$ must extend the longest prefix of all honest validators' bookmarks.
Finally, the validator gossips a ``genesis'' message, which is a vote on the new genesis.
Clients, on receiving the ``recover'' message from the environment, at round $\rrec$, or upon waking after $\rrec$, freeze their ledgers if they haven't done so already (\cref{loc:recovery-client-freeze}).
Clients then wait for ``genesis'' messages and set their new genesis to be one that is included in the ``genesis'' messages from a majority of $\valsetnew$.
Both clients and validators restart the protocol with the new genesis and validator set (\cref{loc:recovery-val-restart,loc:recovery-client-restart}).

During this part, we need to ensure that all honest validators compute the same genesis and that they restart the protocol at the same time.
This can be easily achieved by using any solution to the Byzantine generals problem~\cite{byzantine-generals} that ensures the following properties under honest majority:
\begin{enumerate}
    \item If an honest validator \textsc{broadcast}s a bookmark (\cref{loc:recovery-broadcast-bookmark}), all validators \textsc{deliver} (\cref{loc:recovery-deliver-bookmark}) it at the end of $\bclatency$ rounds.
    \item At the end of $\bclatency$ rounds, all honest validators have \textsc{deliver}ed the same set of bookmarks.
    \item Each honest validator \textsc{deliver}s at most one bookmark per validator.
\end{enumerate}
Solutions to achieve these properties under synchrony are given in~\cite{dolev-strong,byzantine-generals}.
These properties allow all validators to restart their protocol at the same round and with the same genesis.

\section{Analysis}
\label{sec:analysis}

We prove that if $\Pi$ is certifiably safe and live (\cref{def:certifiable-safety,def:liveness}) under honest majority,
the protocol $\PI$ resulting from running the recovery gadget (\cref{alg:protocol}) on $\PI$ 
safely recovers from adversarial majority (\cref{def:rec-liveness}), \ie, it is always safe, live before $\rmaj$, and live soon after $\rrec$.

Toward proving safety, we first show that the bookmarked ledgers of all validators at all rounds before $\rrec$ are consistent with each other.

\begin{lemma}
\label{lem:safety-node-node}
    For all rounds $r_1,r_2 < \rrec$, honest validators $v_1,v_2$, $\LOGacki{v_1}{r_1} \consistent \LOGacki{v_2}{r_2}$.
\end{lemma}
\begin{proof} 
Follows from \cref{lem:freezing-safety} because the validators' code for bookmarking ledgers is the same as the code for confirming ledgers in the freezing gadget (see \cref{alg:protocol}~\cref{loc:recovery-val-code-copy}).
\end{proof}

Second, we prove that the follow-the-leader property (\cref{def:follow-the-leader}) holds until $\rrec$.

\begin{lemma}
\label{lem:safety-node-client}
    For all $r < \rrec$, honest validators $v$ and honest clients $p$, $\LOGfinali{p}{r} \preceq \LOGacki{v}{r}$.
\end{lemma}
\begin{proof} 
Consider an honest client $p$ at let $\LOG = \LOGi{p}{r}$. Then, $p$ saw a \transcript $\transcript$ such that $\untranscribe(\transcript) = \LOG$ by round $r - 3\Delta$. Client $p$ then gossiped $\transcript$ so all validators saw $\transcript$ by round $r - 2\Delta$. Moreover, since $p$ confirmed $\LOG$, we know that $p$ did not see $\transcript'$ such that $\untranscribe(\transcript') = \LOG'$ and $\LOG'\inconsistent \LOG$ until round $r$. This means that no validator saw $\transcript'$ until round $r-\Delta$, because otherwise, the validator would have broadcast $\transcript'$ and client $p$ would have seen $\transcript'$ before round $r$. Therefore, all validators bookmark $\LOG$ or a ledger extending it by round $r$.

\end{proof}

As a corollary of \cref{lem:safety-node-node,lem:safety-node-client}, safety holds for all clients until round $\rrec$.
Due to the follow-the-leader property, we prove that the new genesis computed by honest validators extends the ledgers confirmed by all honest clients before $\rrec$.

\begin{lemma}
\label{lem:recovery}
For all clients $p$ and all round $r < \rrec$, $\LOGfinali{p}{r} \preceq \LOGrec$.
\end{lemma}
\begin{proof}
\newcommand{\LOGcp}{\ensuremath{\LOG_{\mathrm{cp}}}}
    The properties of \textsc{broadcast} and \textsc{deliver} as described in \cref{sec:protocol-recovery} hold.
    Since all honest validators deliver the same set of bookmarks, each validator computes the same $\LOGrec$.
    Let $\LOGcp = \bigcap_{v \in \text{honest}} \LOGacki{v}{\rrec-1}$ be the common prefix of all bookmark ledgers \textsc{broadcast} by honest validators.
    Since each honest validators $v$ \textsc{broadcast}s some $\LOGacki{v}{\rrec-1} \succeq \LOGcp$, due to honest majority among $\valsetnew$, fewer than $|\valsetnew|/2$ \textsc{deliver}ed bookmarks contain any $\LOGacki{}{} \inconsistent \LOGcp$. Hence, $\LOGrec \consistent \LOGcp$ (see \cref{alg:protocol}~\cref{loc:recovery-new-genesis1}). Moreover, $\LOGrec \succeq \LOGcp$ because $\LOGrec$ is the longest ledger whose extensions are bookmarked by $> |\valsetnew|/2$ validators (\cref{alg:protocol}~\cref{loc:recovery-new-genesis}).
    Due to \cref{lem:safety-node-client}, for every client $p$, $\LOGfinali{p}{\rrec-1} \preceq \LOGcp$. Since confirmed ledgers never shrink, for every round $r < \rrec$, $\LOGfinali{p}{r} \preceq \LOGfinali{p}{\rrec-1}$. This concludes the proof.
\end{proof}

This concludes the proof that safety holds at all times (\cref{thm:safety}) since all ledgers confirmed after $\rrec$ will extend all ledgers confirmed before $\rrec$.

\begin{theorem}
\label{thm:safety}
    For any set of rounds $I$, the recoverable protocol $\PI$ is safe in $I$.
\end{theorem}
\begin{proof}
    Due to \cref{lem:safety-node-node,lem:safety-node-client}, for all $r,s < \rrec$ and clients $p,q$, $\LOGfinali{p}{r} \consistent \LOGfinali{q}{s}$.
    Suppose clients $p$ and $q$ restart their protocol (\cref{alg:protocol}~\cref{loc:recovery-client-restart}) at rounds $r_p, r_q \geq \rrec$ respectively.
    Since clients freeze their ledgers at $\rrec$ (\cref{alg:protocol}~\cref{loc:recovery-client-freeze}), for all $r < r_p$, $\LOGfinali{p}{r} \preceq \LOGfinali{p}{\rrec-1}$. Thus, for all $r < r_p, s < r_q$, $\LOGfinali{p}{r} \consistent \LOGfinali{q}{s}$.
    Due to honest majority in $\valsetnew$ and since all honest validators in $\valsetnew$ decide the same $\LOGrec$, clients $p$ and $q$ also decide the same $\LOGrec$ in \cref{alg:protocol}~\cref{loc:recovery-client-genesis}.
    Then, $\LOGfinali{p}{r} \consistent \LOGfinali{q}{s}$ for $r \geq r_p, s \geq r_q$ as well.
    Finally, for $r < r_p, s \geq r_q$, due to \cref{lem:recovery}, $\LOGfinali{p}{r} \preceq \LOGrec \preceq \LOGfinali{q}{s}$, thus $\LOGfinali{p}{r} \consistent \LOGfinali{q}{s}$.
\end{proof}

Liveness before $\rmaj$ and after $\rrec$ follows from the safety and liveness of $\Pi$ during honest majority and the safety of the recovery process (proof in \cref{sec:appendix-proofs}).
\section{Discussion}
\label{sec:discussion}

\noindent
\textbf{Proof-of-Stake.}
We described our gadgets in a permissioned setting for simplicity, but they can be extended to a proof-of-stake setting, where $\valset$ is not a set of validators but a stake distribution. 
Due to our closed-box treatment of the internal protocol $\Pi$, the gadget's working is unaffected by internal updates to the stake distribution by $\Pi$ (\cref{alg:protocol} does not use $\valset$ except during recovery).
During the moment of recovery, analogous to the permissioned setting, the environment announces a new stake distribution $\valsetnew$.
In addition to proof-of-stake implementations of Hotstuff, Casper, and Tendermint, proof-of-stake longest chain protocols~\cite{kiayias2017ouroboros,david2018ouroboros,snowwhite} can also be made certifiable by defining a new confirmation rule for clients~\cite{roughgarden}. Thus, our gadget applies to all these protocols.

\noindent
\textbf{Partial Synchrony.}
Consider partially-synchronous PBFT-style protocols (\eg, Casper~\cite{casper}, Tendermint~\cite{tendermint}, HotStuff~\cite{hotstuff}).
As long as the adversary corrupts less than one-third of the validators, these protocols are live during periods of synchrony and safe even during asynchrony.
It is easy to see that applying our recovery gadget to such a protocol does not harm the above guarantees.
When the adversary corrupts more than one-third validators, if the network is synchronous, our gadget maintains safety and allows for safe recovery of liveness after the adversary falls below $1/3$.
However, because our gadget critically uses synchrony, it does not maintain safety in cases where the network is asynchronous and the adversary exceeds $1/3$ \emph{at the same time}.

\section*{Acknowledgments}
We thank Ertem Nusret Tas, Joachim Neu, Zeta Avarikioti
for discussions related to our recovery model;
Christian Cachin, Giulia Scaffino, and Orfeas Litos
for discussions and reviews of early versions of this paper.
This research was funded by a Research Hub Collaboration agreement with Input Output Global Inc.

The story of how this paper came to be is a curious one. When DZ was completing his PhD in Athens, he had a peculiar encounter with a local taxi driver. It is well known that Greek taxi drivers are ``street smart'' and well versed in a variety of topics. During the drive, he struck up a conversation with the driver, sharing that he is a blockchain researcher. It turns out, the driver had an interest in cryptocurrencies and even owned some Ethereum! The following conversation ensued:

\begin{dialogue}
\speak{Driver} So, tell me, is it true that blockchains break if someone controls more than 50\% of the mining power?

\speak{DZ} Indeed, a 51\% adversary can perform a double spending attack.

\speak{Driver} Well, that's silly. You guys should make it so that a 99\% attacker cannot break the system. Why did you choose 51\%?

\speak{DZ} It is impossible to achieve consensus if the adversary controls more than 51\% of the system's resources, because the honest parties cannot tell each other apart.

\speak{Driver} What a pile of nonsense! Of course you can tell if you are honest.

\speak{DZ} How would you solve the consensus problem for a 99\% adversary then?

\speak{Driver} That's your job to figure out, not mine.
\end{dialogue}

\bibliographystyle{splncs04}
\bibliography{references}

\appendix
\crefalias{section}{appendix}
\crefalias{subsection}{subappendix}
\crefalias{subsubsection}{subsubappendix}

\section{Proof Details}
\label{sec:appendix-proofs}

Since the recoverable protocol involves initializing a new internal protocol after recovery, we define what it means for the internal protocol to be certifiably safe and live during honest majority. This is the property satisfied by existing certifiable protocols such as Sync-Hotstuff~\cite{synchotstuff} and Sync-Streamlet~\cite{streamlet}.

\begin{definition}
\label{def:internal-protocol-security}
    A certifiable protocol $\Pi$ is \emph{certifiably safe and live with latency $u$ under honest majority} if when $\Pi$ is initialized at round $r$ with validator set $\valset$, $\Pi$ is certifiably safe and live with latency $u$ in $[r,r^*)$ where $r^* \geq r$ is the first round in which at least $|\valset|/2$ validators in $\valset$ are corrupted.
\end{definition}

\begin{theorem}
\label{thm:rec-liveness}
    If $\Pi$ is certifiably safe and live with latency $u_{\Pi}$ under honest majority,
    then
    with $u = u_{\Pi} + 3\Delta$
    and $\urec = u_{\Pi} + \bclatency + 4\Delta$,
    for any $\Rhrzn$,
    the recoverable protocol $\PI$ is live with latency $u$ 
    in $[0,\rmaj) \cup (\rrec + \urec, \Rhrzn]$.
\end{theorem}
\begin{proof}
    Consider a round $r$ such that $r < \rmaj$ or $r > \rrec + \urec$. Suppose that all honest validators receive a transaction $\tx$ at round $r' \leq r - u$. 
    Consider a client $p$ that awakens at round $\rboot_{p} \leq r - u$.
    \begin{itemize}
        \item Case 1: $r < \rmaj$. The recoverable client protocol is identical to the freezing protocol before round $\rrec$, except with a $3\Delta$ wait instead of $\Delta$ (\cref{alg:protocol}~\cref{loc:recovery-client-code-copy}). Thus, using \cref{lem:freezing-liveness}, $\PI$ is live with latency $u_{\Pi} + 3\Delta$ in $[0,\rmaj)$.
        \item Case 2: $r > \rrec + \urec$, $r' < \rmaj - u$. Due to Case 1 and \cref{lem:recovery}, $\tx \in \LOGrec$. 
        By round $\rrec + \bclatency$, all honest validators send a ``genesis'' message (\cref{alg:protocol}~\cref{loc:recovery-gossip-genesis}) and by round $\max\{\rrec + \bclatency, \rboot_{p}\} + \Delta < r$, client $p$ receives all honest ``genesis'' messages and thus knows $\LOGrec$ (\cref{alg:protocol}~\cref{loc:recovery-client-genesis}).
        Subsequently, client $p$ restarts its protocol and thus, $\LOGfinali{p}{r} \succeq \LOGrec \ni \tx$.
        \item Case 3: $r > \rrec + \urec$, $r' \geq \rrec + \bclatency$. 
        Note that the instance of $\Pi$ started at $\rrec + \bclatency$ is live after $\rrec+\bclatency$ because of honest majority.
        By round $\max\{\rrec + \bclatency, \rboot_{p}\} + \Delta$, all clients have restarted the protocol, and thus following the argument from Case 1, for $r > \rrec + (\bclatency + \Delta) + u_{\Pi} + 3\Delta$, $\tx \in \LOGfinali{p}{r}$.
        \item Case 4: $r > \rrec + \urec$, $r' \in [\rmaj -u, \rrec + \bclatency)$.
        A transaction sent at such a round $r'$ may not be confirmed by any client or bookmarked by any validator because $\Pi$ may not be live during this period. To ensure such transactions are eventually confirmed, validators must carry over pending transactions that were input before $\rrec + \bclatency$ and consider them as inputs provided to the new instance of $\Pi$ at $\rrec + \bclatency$. Then, following Case 3, $\tx \in \LOGfinali{p}{r}$.
    \end{itemize}
\end{proof}
\section{Comparison with Related Work}
\label{sec:appendix-related}

See \cref{tab:comparison-models-protocols}.

\begin{landscape}
\begin{table}[h]
    \centering
    \caption{Safety and liveness resiliences achievable and impossible in commonly studied models. The safety resilience $\tS$ denotes the maximum adversary fraction up to which the protocol is safe. Similarly, the liveness resilience $\tL$ is the maximum adversary fraction up to which the protocol is live. For each model, we state the pareto-optimal set of resiliences and refer to protocols achieving these resiliences as well as works proving that higher resiliences are impossible.
    From top to bottom, the assumptions in the model weaken and the safety and liveness resiliences decrease.
    We also refer to protocols that recover liveness after a temporary period when the adversary exceeded $\tL$. Note that for rows in which $\tS < 1$, safety may be lost during recovery.}
    \setlength{\tabcolsep}{6pt}
    \setlength{\aboverulesep}{0pt}
    \setlength{\belowrulesep}{0pt}
    \setlength{\extrarowheight}{.75ex}

    {\renewcommand{\arraystretch}{1.5}%

    \begin{tabular}{@{}|p{0.2\columnwidth}|>{\centering\arraybackslash}p{0.1\columnwidth}|>{\centering\arraybackslash}p{0.1\columnwidth}|p{0.15\columnwidth}|p{0.15\columnwidth}|p{0.15\columnwidth}|@{}}%
        \toprule
        \multirow{2}{*}{\textbf{Model}} &
        \multicolumn{2}{c|}{\textbf{Resilience}} &
        \multirow{2}{*}{\textbf{Protocols}} &
        \multirow{2}{*}{\textbf{Impossibility}} &
        \multirow{2}{*}{\textbf{Recovery}} \\
        \cline{2-3} 
          & \textbf{Safety} & \textbf{Liveness} &   &   &   \\
        \hline
        \tikzmark{comparison-most-sychronous}%
        Always-online clients & $\tS = 1$ & $\tL = 1$ & \cite{dolev-strong,bcube,shi-adv-maj,vitalik-dolev-strong} & Not applicable & Trivial \\
        \hline
        Synchrony, client gossip & $\tS = 1 $ & $\tL = \frac{1}{2}$ & This work (\cref{sec:protocol-warmup}) & Future work & This work (\cref{sec:protocol-recovery}) \\
        \hline
        Synchrony, no client gossip & \multicolumn{2}{c|}{$\tL + \tS \leq 1$} & \cite{nakamoto_paper,kiayias2017ouroboros,david2018ouroboros,snowwhite,synchotstuff,streamlet} & \cite{aav1,mtbft} & \cite{zeta-pow-healing,aggelos-pos-healing} \\
        & \multicolumn{2}{c|}{\eg, $\tS = \tL = \frac{1}{2}$} & & & \\
        \hline
        \tikzmark{comparison-least-sychronous}%
        Partial synchrony & \multicolumn{2}{c|}{$2\tL + \tS \leq 1$} & \cite{pbft,hotstuff,streamlet,tendermint,casper,oflex} & \cite{dls88,aav1,oflex} & \cite{zlb} \\
        & \multicolumn{2}{c|}{\eg, $\tS = \tL = \frac{1}{3}$} & & & \\
        \bottomrule
    \end{tabular}

    \begin{tikzpicture}[overlay,remember picture]
        \draw[->] let \p1=(comparison-most-sychronous), \p2=(comparison-least-sychronous) in ($(\x1,\y1)-(0.4,0)$) -- node[label={[rotate=90, anchor=south]left:less synchronous}] {} ($(\x1,\y2)-(0.4,0)$);
    \end{tikzpicture}
    
    }
    
    \label{tab:comparison-models-protocols}
\end{table}
\end{landscape}

\end{document}